\documentclass[fleqn,usenatbib,letters]{mnras}

\usepackage[T1]{fontenc}

\DeclareRobustCommand{\VAN}[3]{#2}
\let\VANthebibliography\thebibliography
\def\thebibliography{\DeclareRobustCommand{\VAN}[3]{##3}\VANthebibliography}

\usepackage{graphicx}	% Including figure files
\usepackage{amsmath}	% Advanced maths commands
\usepackage{subfig}
\usepackage{comment}
\usepackage{float}
\usepackage{enumerate}
\usepackage{newtxtext,newtxmath}
\newcommand{\kms}{\,$\mbox{km s}^{-1}$}

\title[The stellar populations of Mrk 34]{Spatially resolved evidence of the impact of quasar driven outflows on recent star formation : The case of Mrk 34}

\author[P. S. Bessiere et al.]{
P. S. Bessiere$^{1,2}$\thanks{E-mail: patricia.bessiere.astro@gmail.com},
C. Ramos Almeida$^{1,2}$
\\
$^{1}$Instituto de Astrof\' isica de Canarias, Calle V\' ia L\'actea, s/n, E-38205, La Laguna, Tenerife, Spain\\
$^{2}$Departamento de Astrof\' isica, Universidad de La Laguna, E-38206, La Laguna, Tenerife, Spain
}

\date{Accepted XXX. Received YYY; in original form ZZZ}

\pubyear{2021}

\begin{document}
\label{firstpage}
\pagerange{\pageref{firstpage}--\pageref{lastpage}}
\maketitle

\begin{abstract}
We present the results of our spatially resolved investigation into the interplay between the ages of the stellar populations and the kinematics of the warm ionised outflows in the well-studied type II quasar Markarian 34. Utilising IFS data, we determine the spatial distribution of the young stellar population (YSP; $t_{ysp} < 100 \mbox{ Myr}$) using spectral synthesis modelling. We also employ the  $\mbox{[OIII]}\lambda 5007$ emission line as a tracer of the warm ionised gas kinematics. We demonstrate a spatial correlation between the outer edges of the blue-side of the outflow and an enhancement in the proportion of the YSP flux, suggesting that the outflow is responsible for triggering star formation in this region. In regions with more highly disrupted gas kinematics, we find that the proportion of YSP flux is consistent with that found outside the outflow region, suggesting that the increased disruption is preventing a similar enhancement in star formation from occurring. Our analysis suggests that Mrk 34 is an example of quasar driven outflows simultaneously producing both `positive' and `preventive' feedback, further demonstrating the complex nature of the relationship between quasars and their host galaxies.

\end{abstract}


%%%%%%%%%%%%%%%%%%%%%%%%%%%%%%%%%%%%%%%%%%%%%%%%%%

%%%%%%%%%%%%%%%%% BODY OF PAPER %%%%%%%%%%%%%%%%%%

\section{Introduction}
In recent years the role of active galactic nuclei (AGN) in regulating galaxy growth has come to the fore with the advent of cosmological simulations which predict that AGN feedback is required to prevent the growth of over-massive galaxies, thereby reproducing the observed luminosity function \citep{bower06,croton06}. Feedback processes are also credited with mediating the observed relationships between black holes and the properties of their host galaxy's bulge \citep{ferrarese00,gebhardt00,kormendy13}. These feedback processes are usually envisioned as ``negative'', sweeping out and/or heating gas thereby quenching star formation \citep{dimatteo05,hopkins08}.

In contrast, several works (e.g. \citealt{ishibashi12,nayakshin12,silk13,zubovas14,dugan17}) envision a mechanism by which AGN driven outflows produce a ``positive'' feedback effect. In this scenario, the passage of the AGN driven outflow compresses the ambient gas causing gas density to increase at the edges of the bubbles, leading to fragmentation and triggering star formation. In this case, we would expect to see an enhancement of the star formation rate rather than suppression. This has been observed in some active galaxies in both the local and distant Universe (see \citet{cresci18} for a review).

Over the past decade or so, observational studies of gas kinematics in AGN host galaxies have confirmed the presence of energetic outflows in a significant proportion of AGN in both the warm ionised and molecular gas phases(e.g  \citealt{harrison14, cicone14,fiore17, fluetsch19,smethurst21,ramos21}). However, it is less clear what impact these outflows have on the stellar component and whether they are capable of causing the rapid quenching or enhancement in star formation.

A robust approach to understanding the relationship between AGN driven outflows and their impact on star formation is to harness the power of modern integral field spectroscopy (IFS). IFS enables us to simultaneously study the spatial distributions of both stellar populations and AGN outflows, thus uncovering spatial correlations between the two. Indeed, previous such studies (e.g. \citealt{cresci15a, cresci15,carniani16}) have demonstrated negative, preventive and positive feedback in action.

In this study, we endeavour to ascertain whether spatial correlations between the two phenomena exist in the well studied nearby type II quasar Markarian 34. We selected this target, which is part of the QSOFEED sample \citep{ramos19,ramos21}, for an in-depth study because it is known to host an extended, fast outflow \citep{revalski18,trindade21}. Here, we aim to understand any relative differences between the stellar population in regions that are more/less affected by the outflow.

Previous studies of Mrk 34 demonstrate that it is a Compton-thick \citep{gandhi14} quasar at a redshift $z = 0.0501$ with an [OIII] luminosity of $\mbox{L}_{\mbox{[OIII]}} = 10^{8.85} L_{\sun}$ \citep{reyes08}. The quasar is hosted by an Sa-type spiral galaxy which also has a weak bar structure \citep{nair10}, position angle = $30\degr$, inclination $i = 41\degr$, and an ellipticity of $e = 1-b/a = 0.25$ \citep{fischer18}. Deep imaging studies have also shown that Mrk 34 is morphologically disturbed \citep{zhao19,pierce21}, demonstrating recent merger activity. It has a stellar mass  $\log M_* = 11.0 \pm 0.2 M_{\sun}$ \citep{shangguan19} and a star formation rate $\sim 30 \mbox{ M}_{\sun} \mbox{ yr}^{-1}$ \citep{gandhi14}, although the authors caution that this is an upper limit due to the strong contribution of the AGN to the infrared luminosity used to derive this rate. These values place Mrk 34 $\sim 1 \mbox{ dex}$ above the star-formation main sequence. However, in their study of the circum-nuclear regions of a sample of type II quasars, \citet{wang07} found that Mrk 34 falls well below the Kenicutt-Schmidt law, suggesting that it is undergoing suppressed star-formation in the inner-most regions. It is  known to host an energetic and extended outflow with maximum outflow rate of $\dot{M} = 10.3 \mbox{ M}_{\odot} \mbox{ yr}^{-1} $ at 1.26 kpc, a maximum peak kinetic luminosity of $\dot{E} = 1.28 \times 10 ^{43} \mbox{erg s}^{-1}$ at 480 pc and a peak momentum flow rate of $\dot{\rho} = 2.83 \times 10^{34} \mbox{ dyne}$ at 1.32 kpc \citep{trindade21}.

\citet{stoklasova09} also carried out an IFS study of Mrk 34 in which they measured key emission line diagnostics, gas kinematics and made estimations of the mass fractions of the YSPs. However, because they modelled all emission lines as single Gaussian components, they did not capture the detailed outflow kinematics presented here. However, the spatial location of enhanced YSP mass fraction and the higher levels of extinction to the south detected in that work are consistent with the results presented here.

Throughout this paper, we assume $H_0 = 70.0 ~ \mbox{km s}^{-1}~ \mbox{Mpc}^{-1}$, $\Omega_M = 0.3$ and $\Lambda = 0.7$. This results in a cosmologically corrected scale of $1.004 ~ \mbox{kpc arcsec}^{-1}$.

\begin{figure*}
\centering

\subfloat{
\includegraphics[trim=0mm 3.4mm 0mm 1.mm, clip, width=0.33\textwidth]{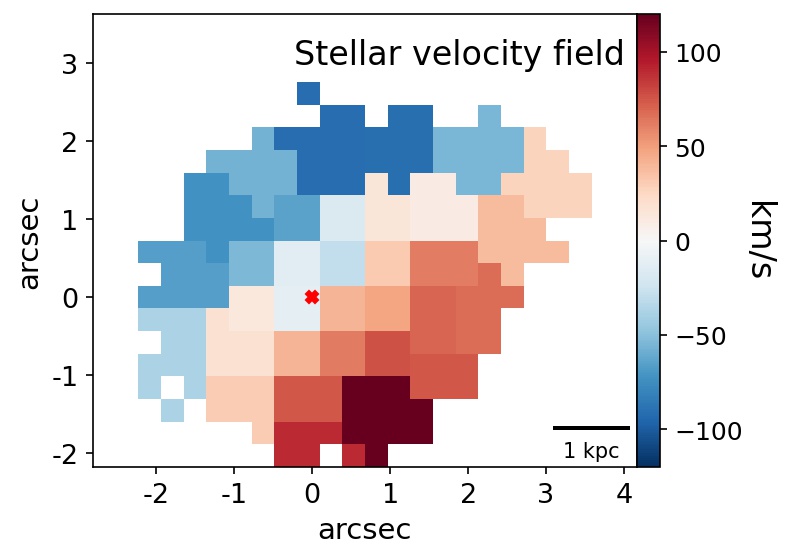}
}
\subfloat{
\includegraphics[trim=0mm 3.4mm 0mm 1.mm, clip,width=0.33\textwidth]{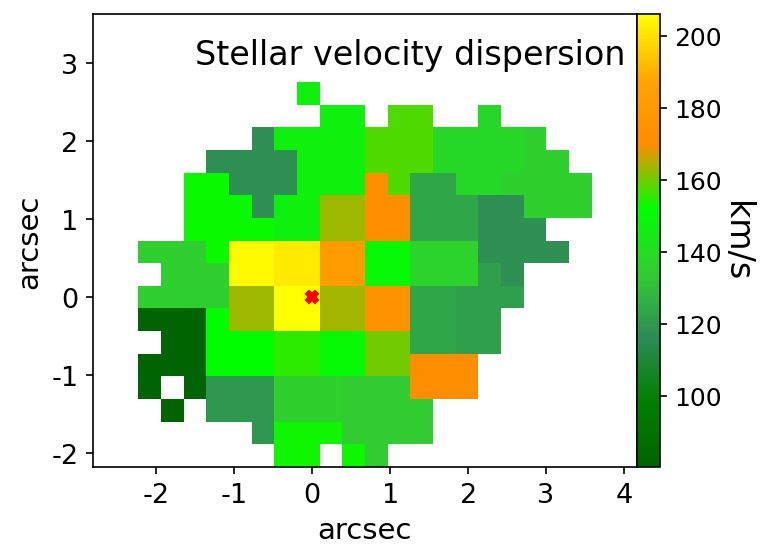}
} 
\subfloat{
\includegraphics[trim=0mm 3.4mm 0mm 1.mm, clip,width=0.33\textwidth]{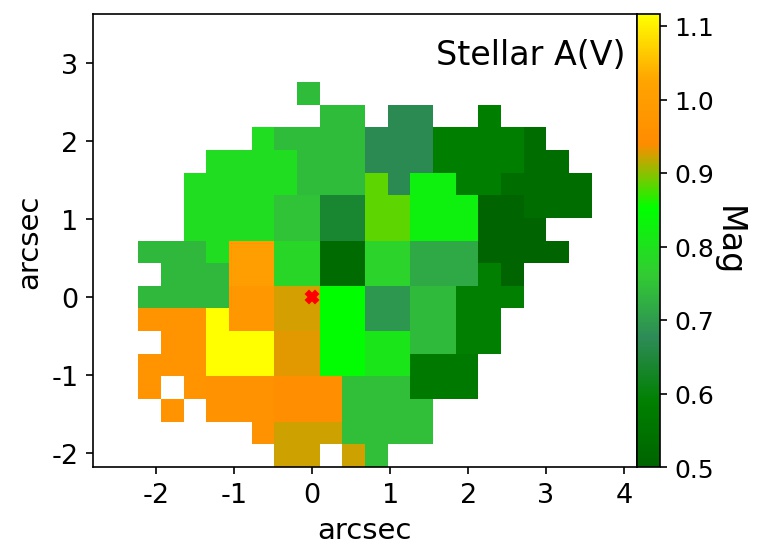}
}

\subfloat{
\includegraphics[trim=0mm 3.4mm 0mm 1mm, clip,width=0.33\textwidth]{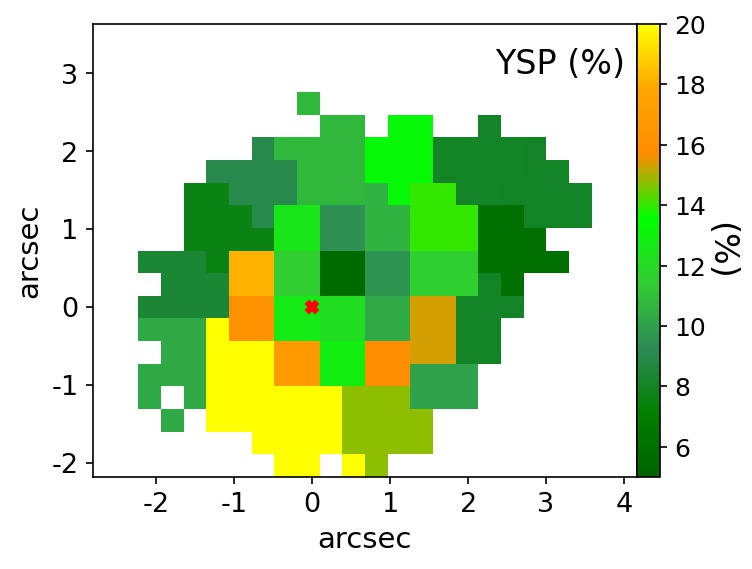}
}
\subfloat{
\includegraphics[trim=0mm 3.4mm 0mm 1mm, clip,width=0.33\textwidth]{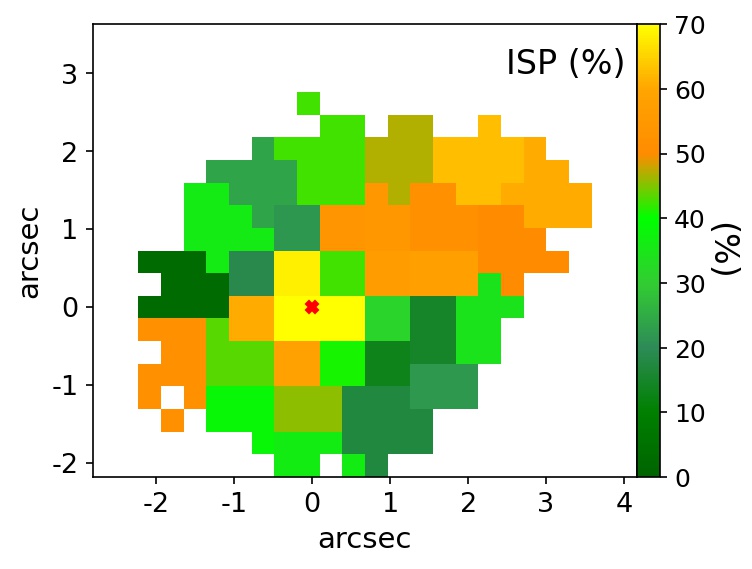}
} 
\subfloat{
\includegraphics[trim=0mm 3.4mm 0mm 1mm, clip,width=0.33\textwidth]{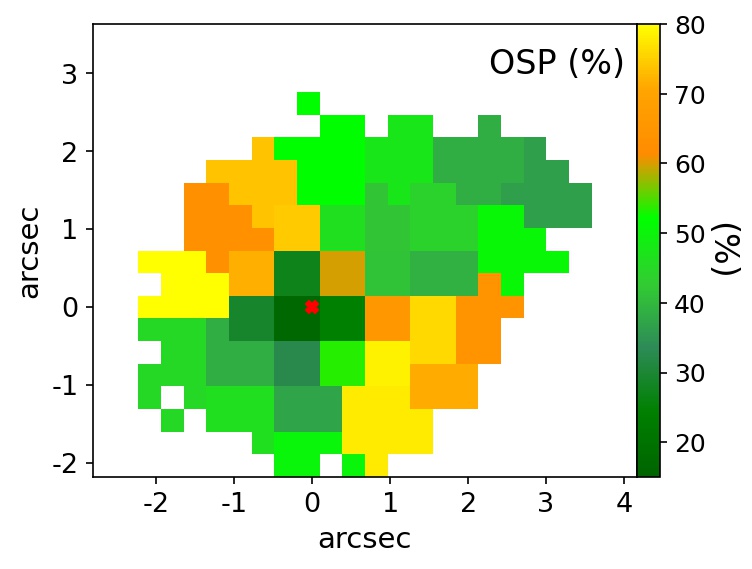}
}

\subfloat{
\includegraphics[trim=0mm 3.4mm 0mm 1mm, clip,width=0.33\textwidth]{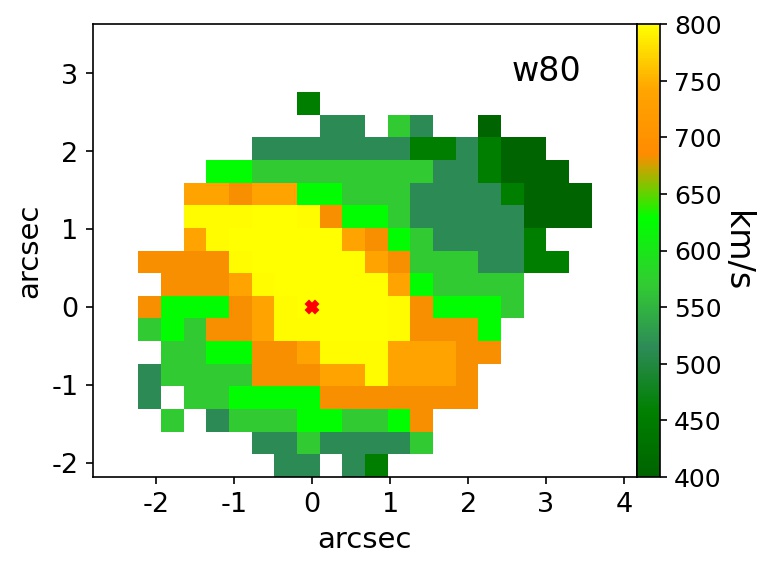}
\label{v80}}
\subfloat{
\includegraphics[trim=0mm 3.4mm 0mm 1mm, clip,width=0.33\textwidth]{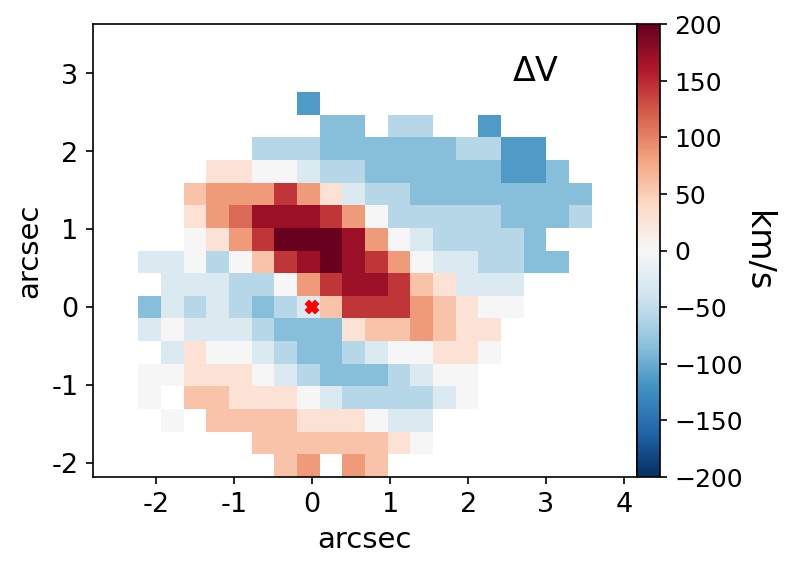}
\label{v90}}
\subfloat{
\includegraphics[width=0.33\textwidth]{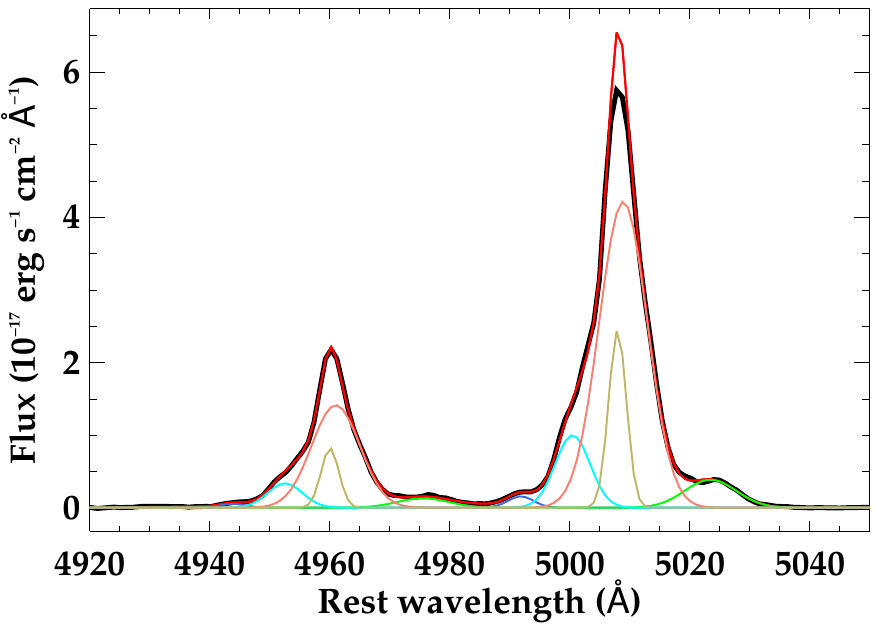}
\label{v96}}

\subfloat{
\includegraphics[trim=0mm 3.4mm 0mm 1mm, clip,width=0.33\textwidth]{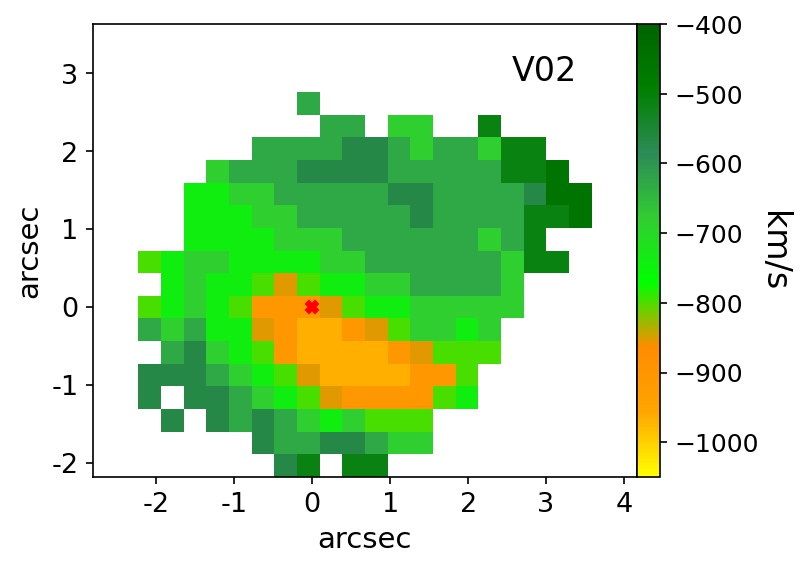}
\label{w80}}
\subfloat{
\includegraphics[trim=0mm 3.4mm 0mm 1mm, clip,width=0.33\textwidth]{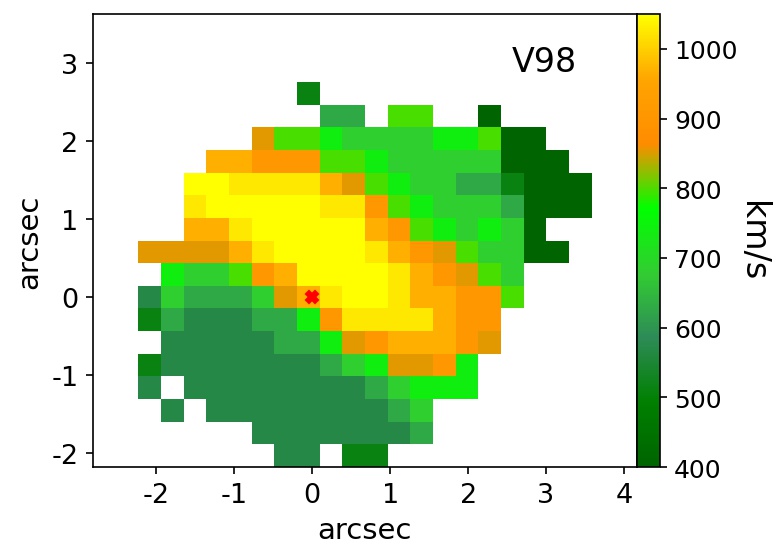}
\label{w90}}
\subfloat{
\includegraphics[width=0.33\textwidth]{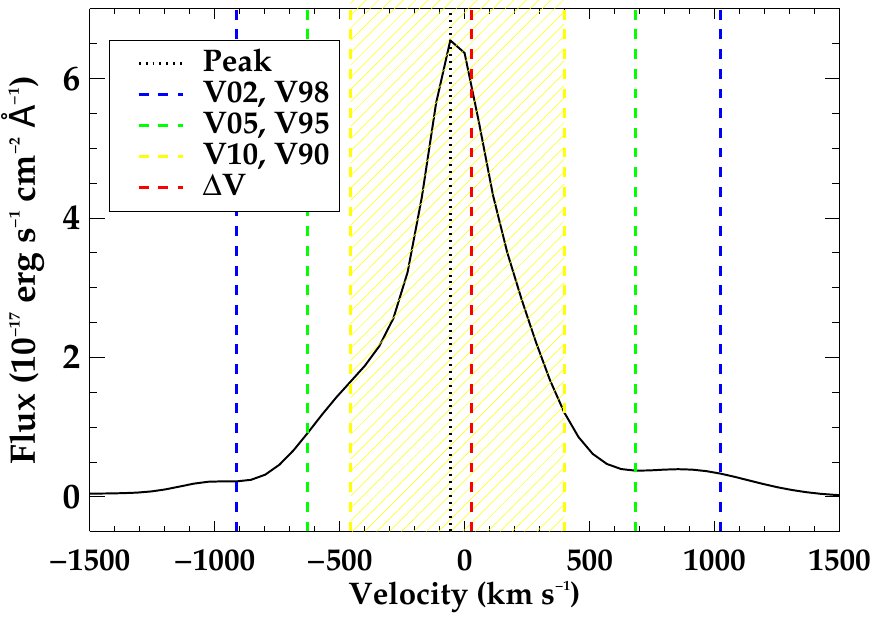}
\label{w96}}

\caption{The maps of the results of the {\sc starlight} modelling and non-parametric kinematic analysis. Top row: stellar velocity field, stellar velocity dispersion and stellar extinction, all derived from the {\sc starlight} modelling. Second row:  Proportion of the total flux in the normalising bin for the YSP, ISP and OSP. Third row: W80 and $\Delta\mbox{V}$ maps derived from non-parametric measurements of the [OIII] lines and an example of a multi component fit to one of the central spaxels. Bottom row: The V02 and V98 maps derived from our non-parametric analysis, representing the maximum gas velocities and an example (same spaxel as above) of the non-parametric method employed in this work. The red cross defines the position of the AGN. North is up and east is to the left.}

\label{non_par}
\end{figure*}

\section{Data reduction and analysis}

These data (retrieved from the Keck Observatory Archive) were obtained using the KCWI IFU mounted on the 10 meter Keck II telescope at Mauna Kea on 06 March 2019 (P.I. A Coil) with both variable seeing and cloud cover throughout the observations (DIMM seeing $\sim 0.6\mbox{\arcsec}-0.9\mbox{\arcsec}$). Mrk  34 was observed for 2800 s ($\sim 45 \mbox{ mins}$) using the BL grating with a medium camera scale resulting in an observed wavelength coverage of 3500 -- 5600\AA\  with a field-of-view of $16.5\arcsec \times 20.4\arcsec $($16.6 \mbox{ kpc} \times \mbox{20.5} \mbox{ kpc}$).  These data were reduced using the IDL version of KDERP data reduction pipeline following the usual steps.

Our main goal is to understand the interplay (if any) between star formation within the central region of the galaxy and the kinematics of the gas that is directly driven by the quasar. Therefore, the region in which we can conduct this study is limited by the signal-to-noise (S/N) in the stellar continuum. In the central region, the spaxels were binned $2 \times 2$ resulting in a bin size of $\approx$0.6\arcsec or 600 pc (approximate seeing at the time of observation) and S/N $> 30 ~ \mbox{pix}^{-1}$ in the normalising bin (4500 -- 4520 \AA\ restframe) used for the stellar population modelling. For the remaining spaxels, we discarded all those with $\mbox{S/N} < 8$ and then employed the Voronoi binning routine of \citet{cappellari03} to bin the spaxels to a minimum S/N = 30, producing a final total of 33 spatial bins. These spectra were then corrected for Galactic extinction \citep{cardelli89} using E(B-V) = 0.008 \citep{schlafly11}, shifted to the restframe and then resampled to 1 \AA $\mbox{ pix}^{-1}$ as recommended for {\sc starlight}.
\subsection{Stellar population modelling}

To carry out the modelling of the spatial bins outlined above, we make use of the stellar synthesis code {\sc starlight
} \citep{fernandes05} in conjunction with the BPASS synthetic stellar population models \citep{eldridge17,stanway18}, which take into account stellar binary evolution. We assume a broken power-law IMF with $\alpha_1 = -1.30$ and $\alpha_2 = -2.35$ and an upper mass cut-off of 100 M$_{\sun}$. We allow for three metallicities accounting for sub-solar, solar and super-solar metallicities (Z = 0.04,0.2 and 0.4 respectively) and use age bins ranging from 0.001 - 2 Gyr in increments of 0.1 dex. Each template represents a simple stellar population formed in a instantaneous burst with an initial mass of $10^6$M$_{\sun}$. 

The modelling was carried in several stages. As the KCWI data does not extend into the red, we first used the SDSS spectrum to determine the composition of the old, underlying stellar population ($t > 2$Gyrs). These components were then combined in the appropriate ratios and used as the input old stellar population for the modelling of each spatial bin. To better fit the stellar populations, the higher-order Balmer emission lines and nebular continuum were modelled and subtracted using the procedure outlined in \citet{bessiere17}. The higher-order Balmer absorption lines were then unmasked and the nebular subtracted spectra were again fit using {\sc starlight}. Throughout the modelling process, we assume a \citet{calzetti00} extinction curve. The results were binned into three groups, young (YSP) < 100 Myrs, intermediate 100 Myr < (ISP) < 2 Gyr and old (OSP) > 2 Gyr. 
\subsection{Emission line fitting}

The kinematics of the warm ionised gas were measured on a spaxel by spaxel basis across the region as outlined above. In order to do so, we adopt a non-parametric approach \citep{harrison14,zakamska14} to measuring the properties of the $\mbox{[OIII]}\lambda 5007$ emission line to fully characterise the kinematics. We first define the values V05,V10,V90 and V95, which measure the velocities at the 5, 10, 90 and 95 per cent points of the normalised cumulative function of the emission line flux. W80 is then a measure of the width of the line containing 80\% of the total flux (W80 = V90 -- V10), whilst $\Delta\mbox{V}$ measures the velocity offset of the broad wings of the emission line from systemic ($\Delta \mbox{V} = (\mbox{V}05 + \mbox{V}95)/2$) ; see \citet{zakamska14} for a detailed explanation). Due to the complex kinematics of the emission lines in some regions, we found that these measures alone did not adequately capture the extent of the velocities of the the outflowing gas. Therefore, we further define V02 and V98, which measure the velocities at the 2\% and 98\% points in the function. These can be considered as the maximum outflow velocity, and in the case of Mrk 34, better captures all the information on the gas kinematics encoded in the emission line (see Figure \ref{non_par}).

The complex gas kinematics in the central region of Mrk 34 results in the blending of the [OIII] doublet. In addition, some spaxels suffer from saturation in the peak of the [OIII]$\lambda5007$ line. To overcome these issues, we first used the {\sc idl} non-linear least-squares fitting routine {\sc mpfit} \citep{markwardt09} to simultaneously fit the  [OIII]$\lambda\lambda4959,5007$ lines with Gaussian profiles, adopting the standard technique of fixing the lines to share the same kinematic components as well as fixing their flux ratio. Due to the analysis method adopted here, we ascribe no physical meaning to individual components and are solely concerned with the goodness of the overall fit. To best achieve this, we begin by fitting two Gaussian components to each line, adding further components (up to a maximum of six in total) until the improvement in reduced $\chi^2 < 10\%$. In cases where $\mbox{[OIII]}\lambda 5007$ was saturated, the peak of the line was masked out resulting in the fit to the peak being driven by $\mbox{[OIII]}\lambda 4959$ line. The resulting $\mbox{[OIII]}\lambda 5007$ line model, comprising the sum of all Gaussian components, is then used as the basis for the non-parametric measurements outlined above. Taking this approach enables us to disentangle the flux from the blended wings of the lines and overcomes the issue of saturation. An example of this process is shown in Figure \ref{non_par} (right column, bottom two panels), where the upper panel shows the fit to the data obtained using {\sc mpfit}, with the data shown in black and the fit shown in red. The colours of the individual Gaussian component is the same in each emission line. The lower panel shows the non-parametric fit to the model shown above. Here, the model is in black, with W80 shown by the yellow shaded region. 

\section{Results}
The results of the procedures outlined above can be seen in Figure \ref{non_par}. The top row shows the velocity field, velocity dispersion and extinction of the stellar component, directly output by {\sc starlight}, for each of the spatial bins. The red point shows the peak of the continuum in the normalising bin, which we take to be the centre of the system and the position of the AGN. The left-hand panel shows the orderly rotation of the galaxy disc, whilst in the right-hand panel, we see that the stellar populations are subject to higher extinction in the SE of the system (A(V) $\approx 1 \mbox{ mag}$) compared to the NW (A(V)$\approx 0.5 \mbox{ mag}$). Assuming that the spiral arms of the galaxy are trailing, the disc is rotating counterclockwise. In conjunction with the stellar velocity field shown in the top left panel of Figure \ref{non_par}, this implies that the NW is the far side of the galaxy. This is consistent with the higher reddening in the SE (i.e. the near side), and is in agreement with the findings of \citet{fischer13}.

%Based on this and the analysis of the morphology of the gas response (i.e. trailing spiral arms), we assume that the NW is the galaxy's far side, in agreement with the findings of \citet{fischer13}.

The second row shows the percentage of the total flux in the normalising wavelength bin for each of the spatial bins. The contribution each population makes varies substantially across the plane of the disc, with the OSP, ISP and YSP accounting for between $\sim$ 16 -- 95 \%, $\sim$ 3 -- 76 \% and $\sim$ 5 -- 25 \% respectively. The maps also show significant segregation of the distributions of the three populations, with the ISP dominating through the central region on a NW to SE axis, whilst the OSP dominates the flux to the SE and NW. Although the YSP never dominates, there is a clear gradient in the proportion of YSP flux across the disc, peaking at $\sim 1.3\arcsec$ or $\sim 1.3 \mbox{ kpc}$ to the SE of the nucleus, which is indicative of an elevated rate of star formation in this region. The coincident increase in stellar extinction also supports the idea that this region is more actively star-forming. It is this \emph{relative} enhancement in star formation that is of particular interest in this work.

The last two rows of Figure \ref{non_par} show the results of our emission line modelling. Here we show the maps of W80, $\Delta\mbox{V}$, V02 and V98.  These maps clearly show the highly disrupted kinematics of the ionised gas within this system. The W80 map demonstrates that across the central region we see velocity widths of the `broad' component ranging from 700 \kms\ to $> 1000$ \kms, indicative of highly disrupted gas. In the map of $\Delta$V, we see a clear double-lobed structure with the blue-side of this highly disturbed gas to the south, reaching a maximum velocity -85 \kms\ , whilst the red-side of the outflow is towards the north and reaches a maximum velocity of 200 \kms. The fact that we detect both the approaching and receding sides of the outflow implies that it is mostly co-planar with the galaxy disc because there is no significant enhancement in attenuation on either side of the outflow (i.e because one lobe is behind the disc). The map of $\Delta\mbox{V}$ in  Figure \ref{non_par} shows that the approaching side of the outflow is in the south, which is also the near-side of the disc. If the outflow were perpendicular to the galaxy disc, the projected outflow velocities would produce a reverse velocity pattern, i.e. blueshifted to the North. Therefore, Mrk 34 is an example of an ionised outflow co-planar with the galaxy disc (i.e maximum coupling between the outflow and ambient gas).

The first two panels of the last row show the maps of V02 and V98 which represent the maximum outflow velocity of the gas. The V02 map shows that the out-flowing gas reaches a maximum projected velocity V02 $\sim -970$ \kms\ towards the south at projected distances of $<$0.5\arcsec directly south of the core and up to 1.5\arcsec to the SW. The V98 map shows that the out-flowing gas reaches a maximum projected velocity V98 $\sim $ 1025 \kms\ starting to the north of the nucleus and extending to around 1\arcsec~projected distance. We also see a region with a slightly lower projected velocity of V98 $\sim$ 1080 \kms\ extending from $\sim$ 1 \arcsec to the SW to $\sim$ 2\arcsec to the NE. These higher V98 velocities (compared to V02) are partly driven by the high-velocity component evident in the third panel of the right-hand column of Figure \ref{non_par} which is redshifted by $\sim 1000 \mbox{\kms}$ from systemic (shown in green).  These findings are consistent with a bi-conical outflow structure embedded in the plane of the disk.

\section{Discussion and conclusions}

\begin{figure*}
    \centering
    \includegraphics[width = 0.8\textwidth]{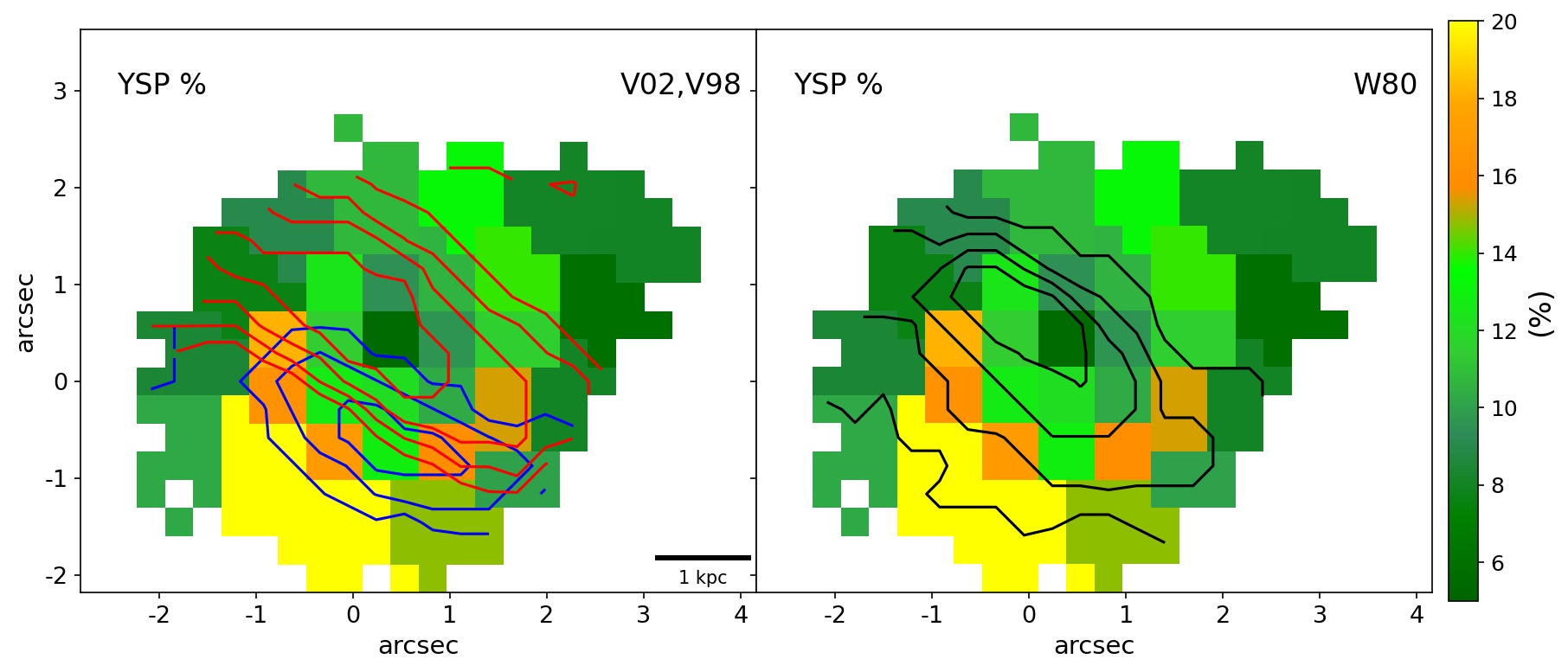}
    \caption{The left panel shows the YSP distribution overlaid with contours of V02 (blue) with contour levels -950,-850 and -750 \kms and V98 (red) with contour levels 750, 850,950,1050 \kms. The right panel shows the same but with contours of W80 overlaid with levels 600,700,800 and 900 \kms.}
    \label{contours}
\end{figure*}

MRK 34 presents an interesting laboratory in which to test the relationship between AGN driven outflows and star formation.  It is a system that hosts a large-scale outflow that is in the plane of the disc, producing the maximum coupling potential between the outflow and ISM, whilst at the same time, appearing to be undergoing suppressed star-formation in the nuclear region. Is this because the outflow is actively suppressing the star formation?

Figure \ref{contours} once again shows the proportion of YSP flux in the normalising bin, with the left panel overlayed with contours of V02 (blue) and V98 (red), whilst the right panel is overlayed with the W80 contours (velocities given in caption). Comparing the kinematics of the outflowing gas with the distribution of the YSP, a relative enhancement is clearly coincident at the edge of the V02 contours, suggesting that a region of enhanced star-formation is coincident with the outer edge of the outflow. This configuration is strongly reminiscent of the spatial locations of AGN triggered star-formation, relative to the outflow, predicted by simulations \citep{ishibashi12,zubovas14}. In this scenario, the ambient gas at the outer edges is compressed, leading to more favourable conditions for star-formation to occur.  The region in which enhanced star formation is detected is also coincident with a region of increased extinction (see Figure \ref{non_par}), which may imply an elevated level of dust and therefore gas relative to the NW of the disc.  Conversely, in regions of lower extinction and, presumably, gas density on the red-side of the outflow, a similar enhancement in YSP flux is not evident. The extinction associated with the stellar populations, as derived from {\sc starlight}, is $\mbox{A(V)} \approx 0.6 \mbox{ mag}$ in the north, whilst in the regions of elevated star formation,  $\mbox{A(V)} >$ 1 \mbox{ mag}. However, the SE is the near-side of the disc where we would expect to measure higher values of extinction. This may be partly or entirely driving the variation in stellar extinction measured between the north and south of the disc.

It is also instructive to consider the time-scales over which the AGN has the potential to impact its host galaxy, in comparison to the ages of the YSPs that we find from our stellar modelling. In the regions of enhanced YSP flux, the modelling suggests YSP ages $1 < \mbox{Myr}$ < 2 which are found to start at $\sim 0.9\arcsec$ or $\sim 0.9 \mbox{ kpc}$ from the center. Taking the mean value of V02 ($\approx 800$\kms) and deprojecting these values using $\mbox{r}_i = r_o/\cos{\alpha}$ and $v_i = v_o/\sin{\alpha}$, where $r_i \mbox{ and } v_i$ are the intrinsic radius and velocity, $r_o \mbox{ and } v_o$ are the observed radius and velocity and $\alpha = 90 - i$ where \emph{i} is the inclination of the galaxy to our line of sight, we find that $r_0 \approx 1.1 \mbox{kpc}$ and $v_o \approx 1200$\kms. This gives a time-scale of $\sim 1 \mbox{ Myr}$ which is consistent with the ages of the YSPs considered here. This could also be interpreted as evidence that, in this particular case, the two phenomena are associated.

It is also evident from Figure \ref{contours} that we do not see a corresponding enhancement in the YSP at the edges of the red side of the outflow, where the level of YSP flux is consistent with those regions outside of the outflow boundary. We note, however, that the red-side of the outflow is associated with higher values of W80 ($\mbox{W80}_{max,r} = 1025 $\kms) compared to the blue-side ($\mbox{W80}_{max,b} = 857$\kms), which could indicate that the outflow is imparting a greater amount of turbulence into the ISM so that the conditions for the triggering of star formation are not similarly enhanced. This supposition is supported by the work of \citet{trindade21}, who measure the kinetic energy and luminosity profiles, as well as the momentum profile and outflow rate of the ionised gas as a function of radius along the NLR. An inspection of Figures 5-9 in that work shows, if we consider radii > 1.1 kpc (where we start to see the enhancement in YSP), the value of each measured quantity is consistently higher to the north-west when compared with the south-eastern side. Therefore, we interpret this as evidence of the outflow preventing star-formation from being triggered rather than actively suppressing ongoing star-formation \citep{pillepich18}.

The findings presented here are consistent with those of previous IFS studies that claim to detect enhanced star-formation at the outer edges of outflows  (e.g. \citealt{cresci15a,cresci15,carniani16,shin19,perna20}). Here, the co-spatial location of the edges of the blue-side of the outflow and the enhancement of YSP flux, in conjunction with the consistent time-scales for the AGN activity and stellar population ages, lead us to conclude that, in this region, we are likely witnessing positive feedback in action (although a chance alignment cannot be ruled out). The outflow compresses the gas at its outer edges as it moves through the ISM, triggering a burst of star-formation. Conversely, in the regions with more highly disrupted gas kinematics, we find that the YSP flux is consistent with that found outside the outflow region. This suggests that the increased disruption is preventing the same enhancement from occurring at the edges of the red-side of the outflow, leading us to conclude that preventive feedback is in action here. This detailed study adds to the growing pool of evidence that AGN-driven outflows are capable of triggering star-formation as well as preventing it, further enhancing our understanding of the complex interplay between AGN and their host galaxy, which may ultimately drive galaxy evolution.

\section*{Acknowledgements}
The authors would like to thank the anonymous referee for their constructive feedback. This research has made use of the Keck Observatory Archive (KOA), which is operated by the W. M. Keck Observatory and the NASA Exoplanet Science Institute (NExScI), under contract with the National Aeronautics and Space Administration. The authors wish to recognize and acknowledge the very significant cultural role and reverence that the summit of Maunakea has always had within the indigenous Hawaiian community.  We are most fortunate to have the opportunity to conduct observations from this mountain. CRA and PSB acknowledge support from the project ``Feeding and feedback in active galaxies'', with reference
PID2019-106027GB-C42, funded by MICINN-AEI/10.13039/501100011033. 
This publication is part of the project ``Quantifying the impact of quasar feedback on galaxy evolution'', with reference EUR2020-112266, funded by MICINN-AEI/10.13039/501100011033 and the European Union NextGenerationEU/PRTR.

%%%%%%%%%%%%%%%%%%%%%%%%%%%%%%%%%%%%%%%%%%%%%%%%%%
\section*{Data Availability}
These publicly available data were downloaded from the KECK data archive \url{https://koa.ipac.caltech.edu/cgi-bin/KOA/nph-KOAlogin}



\bibliographystyle{mnras}
\bibliography{example} % if your bibtex file is called example.bib

%\begin{thebibliography}{99}
%\end{thebibliography}

%%%%%%%%%%%%%%%%%%%%%%%%%%%%%%%%%%%%%%%%%%%%%%%%%%

%%%%%%%%%%%%%%%%% APPENDICES %%%%%%%%%%%%%%%%%%%%%

%%%%%%%%%%%%%%%%%%%%%%%%%%%%%%%%%%%%%%%%%%%%%%%%%%

% Don't change these lines
\bsp	% typesetting comment
\label{lastpage}
\end{document}